\documentstyle[preprint,aps]{revtex}

\begin{document}

\draft

\title{On the Stark broadening of the B III $2s-2p$ lines}
\author{Hans R. Griem\thanks{%
Varon Visiting Professor, permanent address: Institute for Plasma Research,
University of Maryland, College Park, Maryland 20742} and Yuri V. Ralchenko%
\thanks{%
Electronic mail: fnralch@plasma-gate.weizmann.ac.il}}
\address{Department of Particle Physics, Weizmann Institute of Science, Rehovot\\
76100, Israel}
\author{Igor Bray}
\address{Electronic Structure of Materials Centre, School of Physical Sciences, \\
The Flinders University of South Australia, G.P.O. Box 2100, Adelaide 5001,\\
Australia}
\date{Received ...}
\maketitle

\begin{abstract}
We present a quantum-mechanical calculation of Stark line widths from
electron-ion collisions for the $2s_{1/2}-2p_{1/2,3/2},$ $\lambda =2066$ and 
$2067$ \AA ,\ resonance transitions in B III. The results confirm the
previous quantum-mechanical R-matrix calculations but contradict recent
measurements and semi-classical and some semi-empirical calculations. The
differences between the calculations can be attributed to the dominance of
small $L$ partial waves in the electron-atom scattering, while the large
Stark widths inferred from the measurements would be substantially reduced
if allowance is made for hydrodynamic turbulence from high Reynolds number
flows and the associated Doppler broadening.
\end{abstract}

\pacs{32.70.Jz, 34.80.Kw, 52.55.Ez}

\section{Introduction}

The Stark broadening of spectral lines is due to interactions of the
emitting atom (ion) with electrons and ions in a plasma \cite{Griem74}. The
resulting line profiles can serve as an important tool for plasma
diagnostics in a very broad range of plasma parameters. It should be
emphasized, however, that Stark broadening diagnostics generally require
quite elaborate calculations. Therefore, comparison of theoretical line
profiles with the Stark widths measured for well determined plasma
conditions is very important for the improvement of theoretical
approximations and techniques.

Recently, accurate line profile measurements of the $2s-2p$ fine-structure
components of the resonance doublet in Li-like Boron were performed by
Glenzer and Kunze \cite{GleKun96}. They used a homogeneous plasma region in
a gas-liner pinch discharge; and plasma parameters, such as local electron
density and ion temperature, were independently determined by 90$^{o}$
collective Thomson scattering. The Stark line widths were measured to be $%
w\simeq 0.22$ \AA\ for an electron density $N_{e}=1.8\cdot 10^{18}$ cm$^{-3}$
and temperatures $T_{i}=T_{e}=10.6$ eV. This value of $w$ is within 25\% of
the results of semi-empirical \cite{Griem74,HeyBre79} and semi-classical 
\cite{DSB96,Alexiou97} calculations and exceeds the quantum-mechanical
R-matrix calculations \cite{Seaton88} almost by a factor of 2. A similar
discrepancy between measurements and quantum-mechanical calculations had
been noticed previously for $2s-2p$ resonance transitions in another Li-like
ion, namely, Be II \cite{SBJ73}.

In this paper we calculate the Stark line width of the $2s-2p$\ transitions
in B III. In Sec. II the theoretical approach and atomic data used in our
calculations are presented. Then, in Sec. III, we present the results and
discuss the reasons for differences between quantum-mechanical and other
calculations as well as experiment. Finally, Sec. IV contains conclusions
and suggestions.

\section{Theory}

As was shown by Baranger \cite{Bar583}, for an isolated line corresponding
to a transition $u\rightarrow l$ the full collisional width at half-maximum
(FWHM) is given by:

\begin{equation}
w=N_{e}\int_{0}^{\infty }vF\left( v\right) \left( \sum_{u^{\prime }\neq
u}\sigma _{uu^{^{\prime }}}\left( v\right) +\sum_{l^{\prime }\neq l}\sigma
_{ll^{^{\prime }}}\left( v\right) +\int \left| f_{u}\left( \theta ,v\right)
-f_{l}\left( \theta ,v\right) \right| ^{2}d\Omega \right) dv,  \label{eq1}
\end{equation}
where $N_{e}$ is the electron density, $v$ is the velocity of the scattering
electron, and $F\left( v\right) $ is the Maxwellian electron velocity
distribution. The electron impact cross sections $\sigma _{uu^{^{\prime
}}}\left( \sigma _{ll^{^{\prime }}}\right) $ represent contributions from
transitions connecting the upper (lower) level with other perturbing levels
(indicated by primes). In Eq. (\ref{eq1}), the $f_{u}\left( \theta ,v\right) 
$ and $f_{l}\left( \theta ,v\right) $ are elastic scattering amplitudes for
the target ion in the upper and lower states, respectively, and the integral
is performed over the scattering angle $\theta $, with $d\Omega $ being the
element of solid angle. Equation (\ref{eq1}) relates a line width in the
impact approximation with atomic cross sections, facilitating the use of
well-developed techniques of atomic scattering calculations for line
broadening studies. The inelastic terms account for broadening due to
life-time shortening, i.e., broadening associated with decaying amplitudes
of the emitted waves. The elastic terms are due to phase shifts between wave
trains before and after collisions; these phase shifts arise from the
differences in perturbations of upper and lower levels.

The electron impact broadening of the $2s-2p$ principal resonance lines in
Li-like ions differs from the broadening of other lines (like, e.g., $3s-3p$%
) due to the specific level structure of Li-like ions. Both initial and
final levels of this transition are well separated from the other excited
levels, the energy difference between $2s$ and $2p$ states being much
smaller than the energy gap to the nearest $n=3$ level [in B III $\Delta
E\left( 2s-2p\right) \approx 6.0$ eV, while $\Delta E\left( 2p-3s\right)
\approx 16.3$ eV]. Hence, the $\Delta n\geq 1$ inelastic collisions are only
marginally important for the broadening of this line. Additionally, the
temperature of maximal abundance of B III in plasmas is a few times smaller
than its ionization potential $37.9$ eV unless the plasma is rapidly
ionizing. Therefore, it is hardly possible to have B III resonance lines in
high-temperature, high-density plasmas where inelastic perturbations due to
interactions with $n\geq 3$ levels would become important.

Quite surprisingly, practically no accurate atomic data were available for B
III until very recently. The evaluated bibliographic compilation of electron
impact excitation cross sections for ions \cite{PrGal92} contains only {\em %
one} paper on B III with poor accuracy. This differs drastically from the
other members of the Li-like sequence (Be II, C IV, etc.), where on the
average more than 15 papers were published for each ion, some calculations
being claimed to be accurate to within $10\%$. Fortunately, very accurate
results for excitation cross sections from the ground state of B III have
recently been achieved \cite{Marchalant1997}. They were obtained with two
new methods in atomic collision theory, viz., Convergent Close-Coupling
(CCC) \cite{Bray94,BrSt95} and R-Matrix with pseudostates (RMPS) \cite
{Bart96}, which proved to be very successful in calculations of electron
scattering on quasi-one-electron ions (see, e.g. \cite{BartBr97}). Although
the CCC and RMPS methods are quite close in principle, the agreement of {\em %
independently} obtained results which are separately checked for convergence
in coordinate and momentum subspaces is a very convincing argument for these
data to be accurate. The CCC method is a standard close-coupling approach
where all target states (discrete and continuum) are obtained by
diagonalizing the Hamiltonian in a large orthogonal Laguerre basis, and the
coupled equations are formulated in momentum space. Therefore, the
convergence can be easily tested by simply increasing the basis size. The
use of momentum space allows one to avoid the common difficulties related to
the oscillating behavior of wavefunctions in coordinate space. The RMPS
method \cite{Bart96} is a modification of the standard low-energy R-matrix
approach \cite{BurBer93}, where a much larger number of pseudo-orbitals is
taken into account. This significantly improves the description of both the
physical target states and highly excited and continuum pseudostates. For
details on these methods see, e.g., \cite{Bart96,FurBr97}.

For calculations of $2l-3l^{\prime }$ and $2l-4l^{\prime \prime }$ cross
sections, which are relatively small, we used the Coulomb-Born-exchange
(CBE) code {\em ATOM} (the details of the basic approximations can be found
in \cite{SheVa93}). It should be noted that in addition to accounting for
Coulomb attraction and exchange effects, {\em ATOM} calculates inelastic
cross sections with {\em experimental} energy differences between the states
involved and allows for normalization (unitarity) effects. Unlike more
sophisticated and time-consuming CCC and RMPS codes running at least on
workstations, {\em ATOM} quickly generates many cross sections on a modest
PC, which makes it especially suitable for large-scale collisional
calculations. Although the application of the CBE method to a relatively low
charge ion such as B III may be questioned, the difference between the {\em %
ATOM} and CCC/RMPS cross sections is mostly only about 30\% at threshold.
(For highly charged Li-like ions the CBE and CCC results agree much better
with each other \cite{FiRaBe97}.)

\subsection{$2s$ and $2p$ effects}

\subsubsection{Elastic collisions}

The non-Coulomb elastic scattering amplitudes $f_{2s}\left( \theta ,v\right) 
$ and $f_{2p}\left( \theta ,v\right) $ were calculated with the CCC method.
The corresponding elastic cross sections $\sigma _{2s}(E)$ and $\sigma
_{2p}(E)$ as well as the elastic difference term $\tilde{\sigma}(E)\equiv
\int \left| f_{2s}\left( \theta ,v\right) -f_{2p}\left( \theta ,v\right)
\right| ^{2}d\Omega $ are presented in Fig. 1 as a function of the electron
energy in the range $E=0.2-21$ eV. One can see a noticeable difference in
the energy dependence of these parameters. While $\sigma _{2s}(E)$ and $%
\sigma _{2p}(E)$ approximately behave as $1/E$, the elastic term $\tilde{%
\sigma}(E)$ decreases much faster, so that for {\em this} energy region it
can be well fitted by the function $1/E^{\alpha }$ with $\alpha \simeq 1.8$
[for the smallest energies $\tilde{\sigma}(E)$ $\sim \sigma _{2s}(E)$ since,
as is seen from Fig. 1, $\sigma _{2p}(E)$ $\ll \sigma _{2s}(E)$ at $%
E\rightarrow 0$]. The contribution of the elastic term to the line width at $%
N_{e}=1.8\cdot 10^{18}$ cm$^{-3}$ and $T_{e}=10.6$ eV is $w_{el}\approx 0.035
$ \AA , whereas simply using the sum of the elastic cross sections would
give $w_{el}\approx 0.20$ \AA . Due to this cancellation, there may
therefore be more uncertainty in $w_{el}$ than in the following calculation
of inelastic contributions.

\subsubsection{Inelastic collisions}

The $2s-2p$ excitation cross section calculated by various methods is shown
in Fig. 2 as a function of the incoming electron energy. One can see that
contributions of resonances to this cross section are very moderate. It
should be noted also that even at threshold the electron exchange
contributes not more than 10\% of the total cross section. For small
energies the CBE cross sections lie systematically above the RMPS result,
but the difference is only 20\% near the energy of interest. As the
semi-empirical Van Regemorter formula for excitation cross sections (see,
e.g., \cite{PrGal92})

\begin{equation}
\sigma _{se}\left( E\right) =\pi a_{0}^{2}\ f_{ul}\ \frac{8\pi }{\sqrt{3}}%
\frac{Ry^{2}}{\Delta E}\ \frac{\bar{g}(E)}{E}  \label{Gaunt_cs}
\end{equation}
is often used in line broadening calculations, we also show two cross
sections obtained with different choices of the effective Gaunt factor $\bar{%
g}(E)$. In Eq. (\ref{Gaunt_cs}), $a_{0}=0.529\cdot 10^{-8}$ cm is the Bohr
radius, $f_{ul}$ is the absorption oscillator strength, $\Delta E$ is the
energy difference, and $Ry=13.61$ eV. For an absorption oscillator strength
of $f_{2p-2s}=0.365$ and the $\Delta n=0$ Gaunt factor \cite{YouWiese79}

\begin{equation}
\bar{g}(E)=\left( 1-\frac{1}{Z}\right) \left( 0.7+\frac{1}{n_{l}}\right)
\left( 0.6+\frac{\sqrt{3}}{2\pi }\ \ln \left( \frac{E}{\Delta E}\right)
\right) ,  \label{Younger}
\end{equation}
with $Z$ being the spectroscopic charge and $n_{l}$ the principal quantum
number of the lower level, the corresponding cross section is quite accurate
for not too high energies (dot-dashed line on Fig. 2), while the Gaunt factor

\begin{equation}
\bar{g}(E)\simeq 0.8+\frac{\sqrt{3}}{2\pi }\ \ln \left( \frac{E}{\Delta E}%
\right)  \label{Samps}
\end{equation}
recommended in Ref. \cite{SamZha92} (dashed line on Fig. 2) leads at low
energies to an overestimation of the cross section by a factor of 2. This
clearly demonstrates that one should be very cautious when choosing a
specific form of the Gaunt factor. (Measurements of near threshold cross
sections and excitation rate coefficients for C\ IV \cite{Savin1995} also
favor Eq. (3) for $\bar{g}$, while earlier plasma measurements \cite{Griem88}
for N V, O VI and Ne VIII at temperatures well above $\Delta E$ give
effective Gaunt factors that are mostly smaller than those according to Eqs.
(\ref{Younger}) and (\ref{Samps}) by factors 1.5 to 2.)

In Fig. 3 we present the ratio of the sum of partial cross sections with
angular momentum $L\leq L_{T}$ to the total $2s-2p$ excitation cross section
at $E=10$ eV calculated with the CCC and CBE methods (here $L_{T}$ is the
total angular momentum of the system ion+electron). Remarkably, both
approximations show very similar behavior for this ratio with slight
deviations for very low $L_{T}$. It is obvious that the major contribution
comes from rather small $L$ partial waves, so that the partial cross
sections with total angular momentum up to $L_{T}=4$ give more than 60\% of
the total cross section. Even smaller $L$ values dominate the elastic
scattering contribution for which monopole ($\Lambda =0$) contributions are
very important.

To summarize, for an electron density $N_{e}=1.8\cdot 10^{18}$ cm$^{-3}$ and
an electron temperature $T_{e}=10.6$ eV the contribution of the $2s-2p$
inelastic transitions, i.e. of excitation and de-excitation, to the line
width is $w_{in}\left( \Delta n=0\right) \approx 0.062$ \AA . The CBE method
gives $w_{in}\left( \Delta n=0\right) \approx 0.075$ \AA ,\ which is only
about 20\% higher.

\subsection{Inelastic $\Delta n\geq 1$ collisions}

The inelastic cross sections for the transitions $2l-3l^{\prime }$ and $%
2l-4l^{\prime \prime }$ were calculated with the CBE code {\em ATOM} with no
resonances included. Although resonances in the excitation cross sections
are more important for $\Delta n\neq 0$ transitions than for $2s-2p$ \cite
{Marchalant1997}, the contribution of the $2-3$ and $2-4$ inelastic channels
to the total line width is, in fact, rather small. Besides, the comparison
of the RMPS and CBE $2s-3l$ excitation cross sections shows that the
inelastic rate coefficients $\left\langle \sigma v\right\rangle _{2-3}$
produced with {\em ATOM} differ by about 20\% on the average. This accuracy
seems to be quite acceptable since the contribution of $2l-3l^{\prime }$
inelastic transitions to the Stark line width is only $w_{in}\left( \Delta
n=1\right) \approx 0.005$ \AA . Finally, the contribution of the $%
2l-4l^{\prime \prime }$ transitions is one order of magnitude smaller.

Generally, in addition to the electron impact excitation and de-excitation,
other processes of plasma particle scattering from the upper and lower
levels should be taken into account as well. Our CBE estimates and the
CCC/RMPS data \cite{Marchalant1997} show that for the $2s-2p$ line electron
impact ionization and recombination can be safely neglected for the plasma
parameters of Ref. \cite{GleKun96}. Recent semi-classical results \cite
{DSB96} indicate that ion-ion collisions may contribute up to 10 \% to the
total Stark width, but most calculations cited below do not take this effect
into account.

\section{Discussions and Suggestions}

The sum of all electron collisional contributions to the FWHM calculated
here is:

\begin{equation}
w=w_{el}+w_{in}\left( \Delta n=0\right) +w_{in}\left( \Delta n\neq 0\right)
\approx 0.104\text{\AA }  \label{fwhm}
\end{equation}
for an electron temperature of $T_{e}=10.6$ eV and an electron density of $%
N_{e}=1.8\cdot 10^{18}$ cm$^{-3}$. In Table I we compare the experimental
line width $w_{\exp }$ \cite{GleKun96} with different theoretical
calculations \cite{Griem74,HeyBre79,DSB96,Alexiou97,Seaton88}. The two last
columns in this Table present the results of our calculations, viz., the
next to the last column corresponds to Eq. (\ref{fwhm}), while the line
width in the last column was obtained with CBE for all inelastic cross
sections and the elastic term from CCC. The available B III $2s-2p$ Stark
widths are also shown in Fig. 4 as a function of electron temperature. Our
calculations presented here agree with the R-matrix results of Seaton \cite
{Seaton88} practically for all temperatures, thereby confirming the
discrepancy with experiment. The 20\% difference in CCC and R-matrix line
widths for very small $T_{e}$, where elastic collisions dominate, seems to
be related to the strong cancellation effects in the scattering amplitude
difference (see Sec.II). On the other hand, the latest semi-classical
calculations \cite{DSB96,Alexiou97} do agree with the previous results
obtained with similar methods \cite{Griem74,HeyBre79} as well as with the
measured value of the line width, although for small $T_{e}$ the results of
Dimitrijevi\'{c} and Sahal-Brechot \cite{DSB96} deviate significantly from
the other calculations. We also show modified semi-empirical results \cite
{DK1981} which, although calculated only up to $\sim 7$ eV, are very close
to both sets of quantum-mechanical calculations.

Having confirmed the results of the quantum calculations of Seaton \cite
{Seaton88}, a first question is why impact-parameter, semi-classical \cite
{DSB96,Alexiou97} and closely related semi-empirical \cite{Griem74,HeyBre79}
methods lead to substantially larger widths (closer to the experiment). The
answer is related to the dominant role of collisions corresponding to total
angular momentum quantum numbers $L_{T}\leq 4$ (see Sec. II) of the
(colliding) electron-ion system. This fact is equivalent to saying that the
spread of wave packets constructed in order to represent the colliding
electrons classically is comparable to or larger than relevant impact
parameters. (The ratio of de-Broglie wavelength $\lambda $ and impact
parameter $\rho $ is $\lambda /\rho =2\pi \hbar /m\rho v=2\pi /L$.) The wave
packet spread leads to a reduction of the electron-ion interaction and thus
to a decrease in the ensuing line width. This occurs because the electric
fields causing Stark broadening are generated by local deviations from
plasma charge neutrality, and because these deviations are reduced over
spatial scales of the order of the de-Broglie wavelength. Note also that
even most recent semi-classical calculations \cite{Alexiou1995} explicitly
account only for long-range, dipole ($\Lambda =1,$ $\propto r^{-2}$) and
quadrupole ($\Lambda =2,\propto $ $r^{-3}$) perturbations, although for
collisions within the perturbed-electron radius there is also the
short-range monopole $\Lambda =0$ term which, for example, has asymptotic
matrix elements $\propto e^{-\gamma r}$ for $S-S$ transitions \cite
{Bates1950}. This term is properly allowed for in the quantum calculations
and further smoothes the electron-ion interactions.

To illustrate the relation between relevant electron-ion separations at the
perihelion $r_{\min }$ of the classical (unperturbed) orbits and angular
momentum $L$ corresponding to the impact parameter $\rho $, consider

\begin{equation}
\frac{r_{\min }}{a_{0}}=\frac{1}{2}\frac{L^{2}}{\left[ 1+\left( 
{\displaystyle {L \over \eta }}%
\right) ^{2}\right] ^{\frac{1}{2}}+1},  \label{periratio}
\end{equation}
which follows from Eqs. (116), (117) and (118) of Ref. \cite{Griem74} with
the Coulomb parameter

\begin{equation}
\eta =\frac{2e^{2}}{\hbar v}  \label{eta}
\end{equation}
for doubly-charged ions and electrons of velocity $v$. Our calculations with
the Hartree-Fock code of Cowan \cite{Cowan1981} show that for B III ions in
the $2s$ and $2p$ states, the corresponding bound state mean radii are close
to $1.6\,a_{0}$, while typical Coulomb parameters range from about 2 to 4
(for $E_{e}=10$ eV, $\eta \approx 2.3$). As can be seen from Table II, the
classical orbits indeed penetrate deeply, or, at least, come to within
factor 2 of the bound state orbits for angular momenta found to be most
important in the quantum scattering calculations (see Fig. 3).

The theoretical conclusion that semi-classical calculations overestimate the
electron collisional broadening of the B III $2s-2p$ lines by a factor of
about 2 leaves one with the dilemma of having about the same disagreement
with the experiment \cite{GleKun96}, for which the combined error in Stark
width and electron density measurements was estimated to about 20\%. A
natural suggestion is to reconsider any possible systematic errors, a
probable cause for such errors being hydrodynamic turbulence associated with
plasma flows in the gas-liner, Z-pinch experiment \cite{GleKun96}, similar
to the MHD turbulence invoked \cite{Thornhill94,Rudakov1997} for
interpretations of high-power Z-pinch experiments. This would be analogous
to a more extreme situation encountered in the measurement of C V $1s2s\
-1s2p\ $lines \cite{IglGriem88} which had also been found to be
substantially broader than predicted theoretically. However, in this
theoretically rather similar case of $\Delta n=0$ transitions of He-like
ions, there are both singlet and triplet lines, at rather different
wavelengths of $\lambda $= 3526\ \AA\ or 2271 and 2277\AA ,\ respectively.
The measured widths were proportional to $\lambda $ rather than $\lambda
^{2} $, indicating Doppler rather than Stark broadening to dominate. Since
these widths were larger than expected from thermal Doppler broadening and
from Doppler shifts associated with radial flows, hydrodynamic turbulence in
the laser-produced plasma used was inferred from this excess broadening,
with an effective temperature of 600 eV.

Although radial laser-blowoff and pinch implosion velocities are very
similar in these experiments, both approaching 10$^{7}$ cm/sec, there is, of
course, an essential difference in that the B III measurements were made in
a 50 nsec interval shortly after maximum compression, while the C V
measurements were taken while axial velocities were still about $3\cdot
10^{7}$ cm/sec. However, this distinction may not be very important, because
any turbulence from high Reynolds number flows decays only on a time scale $%
\tau $ of $l/\Delta v$ \cite{LanLif??}, if $l$ is a characteristic length
and $\Delta v$ a typical spatial difference of flow velocities. With $%
l\approx 1$ cm and $\Delta v\approx 10^{6}$ cm/sec one would thus expect $%
\tau \approx 1$ $\mu \sec $, much too long for any turbulence to decay
before the B III measurement interval.

As to typical Reynolds numbers

\begin{equation}
R=\frac{vl}{\nu }  \label{reynolds}
\end{equation}
{\em during} the pinch implosion, we estimate $R\approx 1.5\cdot 10^{4}$ for 
$v=5\cdot 10^{6}$ cm/sec, $l=3$ cm, and a kinematic viscosity \cite{NRL
formulary},

\begin{equation}
\nu =\frac{3\cdot 0.96T^{5/2}}{4e^{4}\left( \pi m_{i}\right) ^{1/2}N_{e}\ln
\Lambda },  \label{nu}
\end{equation}
of 990 cm$^{2}$/sec, at T = 10 eV, N$_{e}=10^{18}$ cm$^{-3}$ and a Coulomb
logarithm \cite{NRL formulary} of 6.1. Since the implosion takes more than 1 
$\mu $sec, developed and saturated turbulence seems therefore unavoidable, $R
$ being larger than critical Reynolds numbers \cite{LanLif??}. Note also
that any magnetic field effects are not likely to reduce the turbulence
significantly, both because fields and plasma are probably fairly well
separated \cite{Davara} and because the corresponding parameter $\omega
_{ci}\tau _{i}$ is well below 1 in any case (here $\omega _{ci}$ is the ion
gyrofrequency and $\tau _{i}$ is the ion-ion collision time). Another
question is the extent to which the turbulence is transported radially
inward or mixed into the test gas region. However, even local Reynolds
numbers are probably above critical over most of this region and significant
Reynolds stresses would be needed to compensate for the reduction in
particle pressure implied below.

The reader may wonder whether the (collective) Thomson scattering
diagnostics \cite{DeSilva,Wrubel} used in the B III experiment \cite
{GleKun96} would not have indicated the existence of hydrodynamic
turbulence. For sufficiently high concentrations of heavier elements in the
hydrogen fill gas, the so-called impurity peak would then indeed indicate a
higher temperature for these ions than for protons (see, e.g.., Fig. 6 of
Ref. \cite{DeSilva}), although the larger width of this peak near peak
compression could just as well be caused by turbulent velocities close to
the thermal velocities of the protons provided the turbulent eddies are 
smaller than the scattering volume. For relatively high impurity
concentrations, radiative energy losses are very important \cite{BaigKunze},
facilitating a more rapid decay of the turbulence than estimated above. This
is consistent with the narrower impurity peaks at later times (see Fig. 5 of
Ref. \cite{DeSilva}), whose widths are consistent with thermal Doppler
broadening at equal temperatures for the various ions.

The B III experiment, on the other hand, was done with very small
concentrations of BF$_{3}$ to avoid self-absorption of the $2s-2p$ lines.
The impurity feature on the scattering spectrum could therefore not be
observed \cite{Glenzer1996}, and at the same time the dissipation of the
turbulent energy would have taken much longer, say, the 1 $\mu $sec plasma
life time quoted in Ref. \cite{GleKun96}. One is therefore left with the
possibility of an alternative interpretation of the scattering spectra,
e.g., of that shown in Fig. 2 of Ref. \cite{Glenzer1996}. Namely, instead of
inferring a temperature of 10 eV, essentially from the width of the proton
feature, assume turbulent rms velocities equal to proton thermal velocities.
This also means lower temperatures for the protons, say, 5 eV. As emphasized
by Glenzer and Kunze \cite{GleKun96}, electron-ion relaxation times are
extremely short, so that we also infer $T_{e}$ $\simeq $ 5 eV. This is a
much more favorable electron temperature for the observation of the B III $%
2s-2p$ lines than 10 eV in this nearly steady state plasma, because at 5 eV
about 20\% of the boron ions are B III, contrasted to less than 1\% at 10
eV. Independent evidence for $T_{e}\lesssim 5$ eV could be provided by the
absence of the $3d-4f$ line at $2077$ \AA , reported in Ref. \cite{GleKun96}%
, whose intensity was evidently $\lesssim 5\%$ of the 2066 \AA\ line. At 10
eV, the relative intensity of the 2077 \AA\ line would be about 1. (Note
that standard LTE relations can be used for this estimate, to within about
10\% for the temperature. Since the $3d-4f$ line may be about 3\AA\ wide,
corresponding line ratio measurements would best be done at reduced spectral
resolution.) Any deviations between shapes of thermal and thermal plus
turbulence scattering spectra would probably be too small to be observable,
leaving its width as the major invariant. Some deviations could, of course,
be indications of non-Gaussian distributions of the turbulent velocity
components.

Besides suggesting investigations of turbulence in the gas-liner pinch,
perhaps along the lines of Ref. \cite{Wrubel} and by use of line pairs with
different sensitivities to Stark and Doppler broadening \cite{IglGriem88},
it remains to be shown that the level of turbulence assumed here would
suffice to obtain agreement of measured line widths with quantum-mechanical
calculations. Note first that there will be no significant change in the
electron density, $N_{e}=1.8\cdot 10^{18}$ cm$^{-3}$, because of the large
value of the scattering parameter $\alpha $, i.e., of $\left( k\lambda
_{D}\right) ^{-1}$, $k$ here being the wavenumber of the electron density
fluctuation responsible for the scattering and $\lambda _{D}$ the Debye
length \cite{DeSilva}. However, the predicted electron-collisional width of,
e.g., Ref. \cite{Seaton88}, is now 0.14 \AA\ because of the reduced electron
temperature, whereas the predicted total Doppler width is increased by a
factor $\left( 11.8/2\right) ^{1/2}$ to 0.125 \AA , $11.8=10.8+1$ standing
for turbulent plus thermal Doppler broadening at a boron/proton mass ratio
of 10.8, 1/2 for the reduction in temperature. With 0.07 and 0.05 \AA\ %
Lorentzian and Gaussian instrumental broadening \cite{Glenzer1996} and 0.02 
\AA\ proton impact broadening \cite{DSB96}, this gives a total line width 
\cite{Hulst1947,Whiting1968} of 0.28 \AA , i.e., more than 90\% of the
measured total width \cite{GleKun96}. There would therefore be agreement
well within combined experimental and theoretical errors relative to the
previous quantum-mechanical calculations and, more marginally, also with the
present calculations, which result in 0.27 \AA , if one allows again for a
0.02 \AA\ contribution from ion-ion collisions, were the degree of
turbulence indeed as high as assumed here. Verification of this assumption
would remove a major obstacle in our quantitative understanding of Stark
broadening of isolated lines from multiply ionized atoms, including possibly
the anomalous scaling of line widths along isoelectronic sequences \cite
{Bottcher1988,Glenzer1992,Glenzer1994a,Wrubel1996b}. In the last experiment,
on $2s3s-2s3p$ line widths of Be-like Neon, improved semi-classical
calculations \cite{Alexiou1995} were found to be consistent with the
measured widths, but judging from the present B III $2s-2p$ comparisons and
given similar ratios of impact parameters and bound state radii, this
agreement may again be spurious.

\section{Conclusions}

A fully quantum-mechanical calculation of the Stark line width for the $%
2s-2p $ line of B III was carried out with the use of the latest atomic data
reflecting the present state-of-the-art in atomic collision theory. Although
the obtained results agree well with the previous quantum R-matrix Stark
widths, the difference with semi-classical and some semi-empirical
calculations as well as with the measured values is of order of 2. This
seems to originate in (i) failure of the non-quantum calculations for small
impact parameters which are most important for the line width in question,
and from (ii) not accounting for the turbulent plasma motion which
significantly affects the determination of Doppler broadening and plasma
temperature. Independent ion line width measurements for plasmas with
well-known parameters, and not subject to significant contributions from
other line broadening mechanisms than Stark broadening, continue to be very
important.

\section{Acknowledgments}

We are grateful to S.Alexiou, R.Arad, S.B\"{u}scher, V.Fisher, H.-J.Kunze,
Y.Maron, M.Seaton and T.Wrubel for valuable discussions and to S.Alexiou for
sending his results prior to publication. This work was supported in part by
the Israeli Academy of Sciences and the Ministry of Sciences and Arts of
Israel. Research of I.Bray is sponsored in part by the Phillips Laboratory,
Air Force Material Command, USAF, under cooperative agreement number
F29601-93-2-0001, and that of H.R.Griem in part by the US National Science
Foundation.

\newpage

\begin{center}
{\bf Figure Captions}
\end{center}

FIG. 1. Non-Coulomb elastic cross sections from the $2s$ (solid line) and $%
2p $ (dashed line) states of B III vs. electron energy $E$. The elastic
difference term $\tilde{\sigma}\left( E\right) $ is shown by the dot-dashed
line.

\medskip

FIG. 2. Excitation cross section for the B III $2s-2p$ transition as a
function of electron energy: ---, R-matrix with pseudostates \cite
{Marchalant1997}; $\cdot $ $\cdot $ $\cdot $, Coulomb-Born-exchange (Sec.
II); -- $\cdot $ --, semi-empirical Van Regemorter cross section according
to Eq.(\ref{Gaunt_cs}) with Gaunt factor after Ref. \cite{YouWiese79}; -- --
--, semi-empirical Van Regemorter cross section with Gaunt factor after Ref. 
\cite{SamZha92}.\strut

\medskip

FIG.3. Ratio of the sum of partial cross sections with angular momentum $%
L\leq L_{T}$ to the total excitation cross section for the BIII $2s-2p$
transition vs. angular momentum $L_{T}$.

\medskip

FIG.4. Stark widths for the B III $2s-2p$ transition vs. electron
temperature for $N_{e}=1.8\cdot 10^{18}$ cm$^{-3}$. Experimental value from 
Ref. \cite{GleKun96}; theory: present
work ------, semi-empirical (Ref.\cite{Griem74}) $\cdot \cdot \cdot \cdot $,
semi-empirical (Ref.\cite{HeyBre79}) -- -- --, semi-classical (Ref.\cite
{DSB96}) --- --- ---, semi-classical (Ref.\cite{Alexiou97}) $\times $,
R-matrix (Ref.\cite{Seaton88}) -- $\cdot $ --, modified semi-empirical (Ref.%
\cite{DK1981}) $\bullet \bullet \bullet $.

\begin{center}
\strut

\newpage {\bf Tables}

\begin{table}[tbp] \centering%
\caption{Ratio of the experimental Stark width [S.Glenzer and H.-J.Kunze, 
Phys.Rev. A {\bf 53}, 2225 (1996)] of the 2s-2p line in B III to
different theoretical widths.\label{TableI}} 
\begin{tabular}{c|c|c|c|c|c|c|c|c}
\hline
${\bf T}_{e}$ (eV) & ${\bf N}_{e}$ (cm$^{-3}$) & \multicolumn{7}{c}{${\bf w}%
_{\exp }{\bf /w}_{theor}$} \\ \hline
10.6 & 1.81$\cdot 10^{18}$ & 1.2$^{a}$ & 1.2$^{b}$ & 1.1$^{c}$ & 1.0$^{d}$ & 
1.8$^{e}$ & 2.1$^{f}$ & 1.9$^{g}$ \\ \hline
\end{tabular}
\end{table}%
\end{center}

$^{a}$Semi-empirical \cite{Griem74}, $^{b}$semi-empirical \cite{HeyBre79}, $%
^{c}$semi-classical \cite{DSB96}, $^{d}$semi-classical \cite{Alexiou97}, $%
^{e}$R-matrix \cite{Seaton88}, $^{f}$CCC method (present work), $^{g}$CBE
method (present work)

\begin{center}
\begin{table}[tbp] \centering%
\caption{Ratio of the electron-ion perihelion to the Bohr
radius for various values of angular momentum $L$ and the Coulomb parameter 
$\eta$ defined in Eq. (\ref{eta}).\label{TableII}} 
\begin{tabular}{c|ccc}
\hline
L%
\mbox{$\backslash$}%
$\eta $ & 2 & 3 & 4 \\ \hline
1 & 0.24 & 0.24 & 0.25 \\ \hline
2 & 0.83 & 0.91 & 0.94 \\ \hline
3 & 1.61 & 1.86 & 2.00 \\ \hline
4 & 2.47 & 3.00 & 3.31
\end{tabular}
\end{table}%
\end{center}

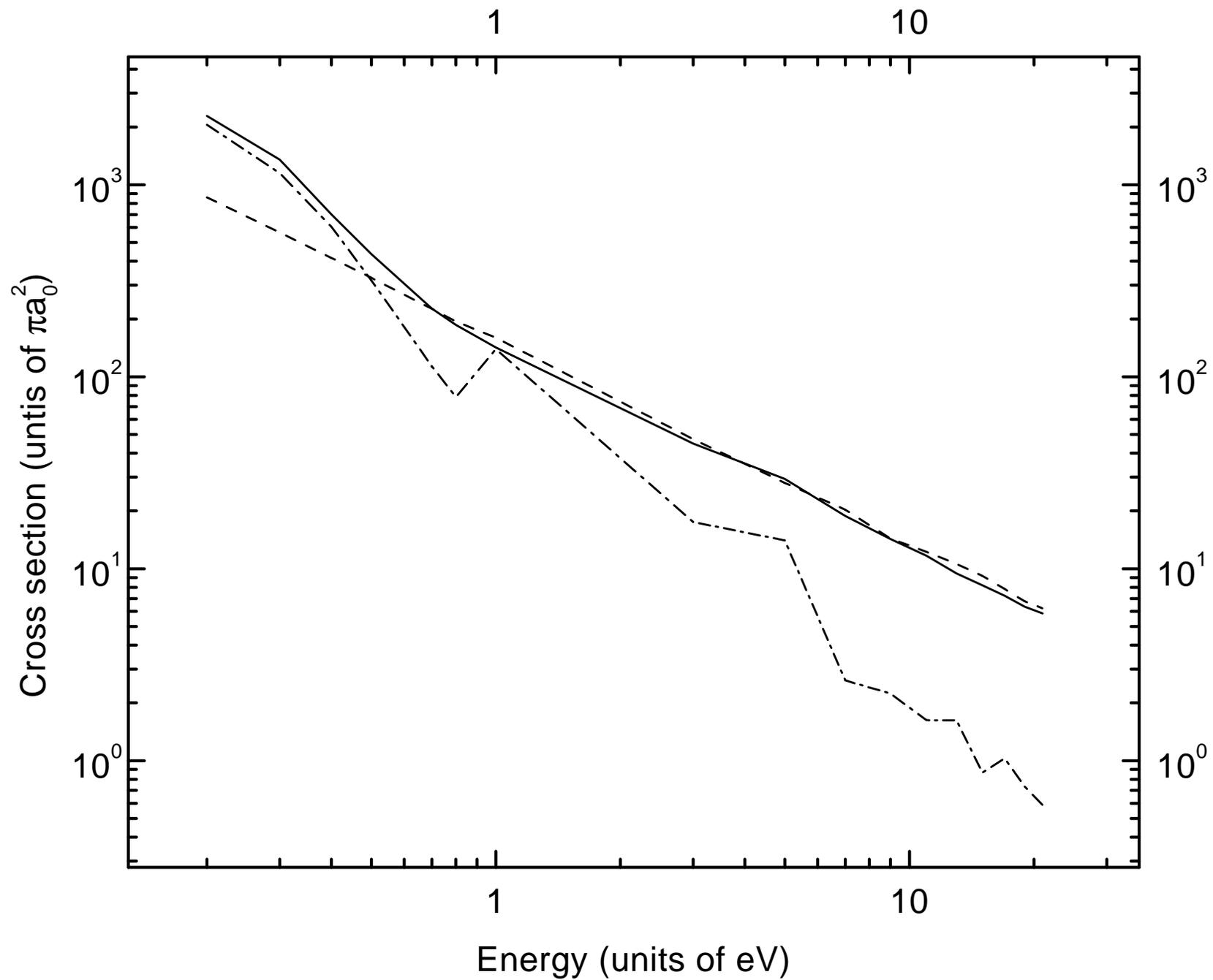
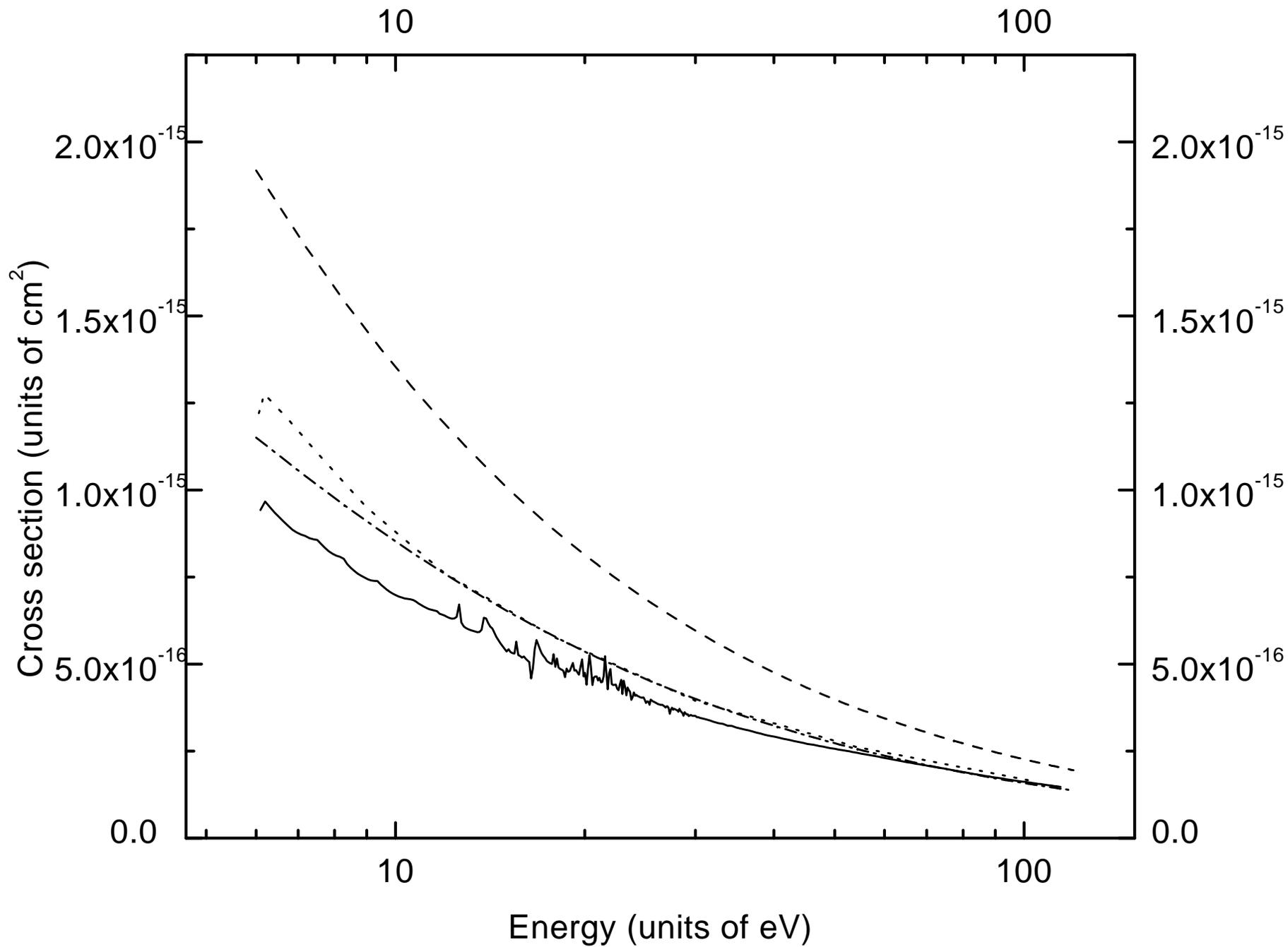
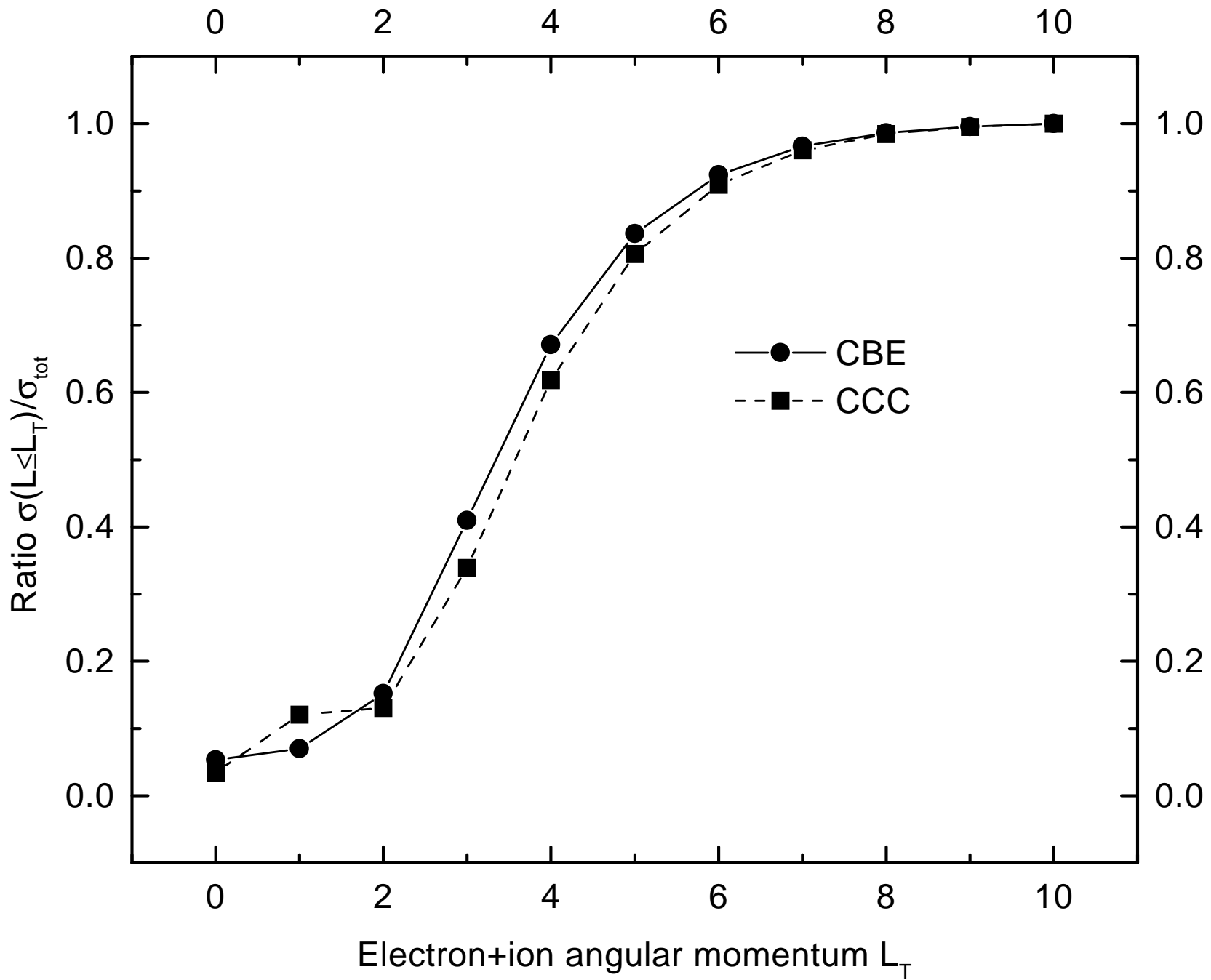
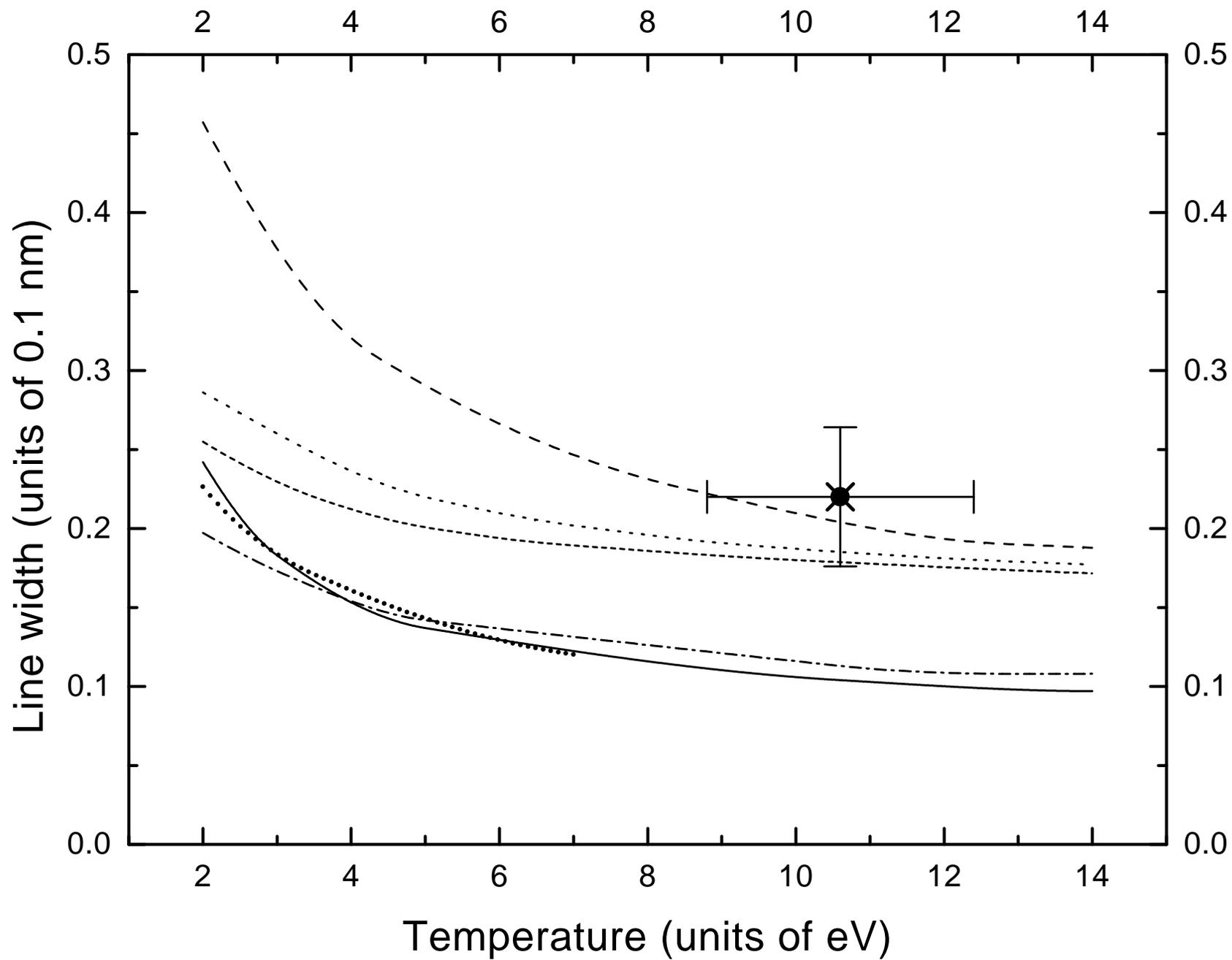
\end{document}